\documentstyle[pra,aps,twocolumn]{revtex}

\begin{document}
\draft
\title{Band Gaps for Atoms in Light based Waveguides}
\author{J.J. Hope and C.M. Savage}
\address{Department of Physics and Theoretical Physics,
The Australian National University, \\
Australian Capital Territory 0200, Australia.}
\date{\today}
\maketitle

\begin{abstract}
The energy spectrum for a system of atoms in a periodic potential can
exhibit a gap in the band structure.  We describe a system in which a
laser is used to produce a mechanical potential for the atoms, and a
standing wave light field is used to shift the atomic levels using the
Autler-Townes effect, which produces a periodic potential.  The band
structure for atoms guided by a hollow optical fiber waveguide is
calculated in three dimensions with quantised external motion.  The
size of the band gap is controlled by the light guided by the fiber.
This variable band structure may allow the construction of devices
which can cool atoms.  The major limitation on this device would be
the spontaneous emission losses.
\end{abstract}


\narrowtext

\section{Introduction} \label{sec:Intro}

Band gaps in the energy spectrum of electrons cause many interesting
effects in solid state physics, and there may be interesting devices
that can be constructed using similar effects in atom optics.  This
paper describes a system that can produce gaps in the energy spectrum,
and calculates its three dimensional energy spectrum.

Standing wave laser beams produce a periodic potential which has been
used extensively in atom optics, particularly in cooling experiments
utilising the Sisyphus effect \cite{Sis2D,Marte,Doery}.  Atoms in a
periodic potential have an energy spectrum consisting of bands
\cite{Wilkens}.  Experiments and theoretical calculations concerning
these systems have been conducted in one and two dimensions.  Three
dimensional optical lattices have been investigated experimentally
\cite{Hemmerich}, but the quantised atomic motion has not been calculated
except in the limit of very deep potential wells\cite{Kastberg}.  We
present a numerical calculation of the energy spectrum of atoms in a
three dimensional waveguide.  This paper shows that the band structure
of these atoms can be manipulated to produce an energy gap between the
lowest bands.

We consider atoms that are strongly confined in two dimensions and
relatively free to travel in a single direction.  Hollow optical
fibers have been suggested as a practicable method of guiding atoms
coherently\cite{Marksteiner,Ito,JILA}.  A laser is guided along the fiber
and the evanescent field in the interior of the waveguide interacts
with the dipole moment of the atoms to form a potential barrier.  A
naive method for producing a periodic potential would be to guide a
standing wave along the fiber, but this allows atoms to absorb a
photon from one beam and emit a photon into the counterpropagating
beam.  The associated recoil is larger than the kinetic energy of the
atoms in the lowest band, so it would destroy any interesting effects.
To avoid this, a second laser beam is introduced into the fiber to
shift the atomic energy levels by the Autler-Townes effect.  This
produces a periodic potential without momentum diffusion.  We present
a detailed calculation of the band structure of atoms guided by a
hollow optical fiber, although our model may be applied to some free
space laser configurations.

In section \ref{sec:Pot} we derive the potential seen by the atoms
under the influence of the two laser fields.  In section \ref{sec:1D}
we present a short revision of the band structure expected for an atom
in a one dimensional sinusoidal potential.  Section \ref{sec:Calc}
describes the calculation of the atomic energy spectrum in the three
dimensional fiber.  The effects of spontaneous emission are estimated
later and are found to be the critical factor which will limit
experimental realisation of such devices.  Section \ref{sec:AltSys}
describes some possible generalizations of this work to other laser
configurations, and section\ref{sec:Appl} describes some applications
of controllable band structure for the atoms in a hollow optical
fiber.  In particular, a technique is described which can cool atoms.

\section{The Potential in a Hollow Optical fiber} \label{sec:Pot}

When the external motion of an atom is treated classically, an atom in
a highly detuned light field will experience an effective potential
that is proportional to the intensity of the field and inversely
proportional to the detuning of the field.  It follows that a standing
wave light field will produce a periodic potential.  When the atoms
are extremely cold, there is a second effect which dominates the
motion.  The standing wave allows atoms to absorb and emit photons in
either direction, which means that the atoms will diffuse in momentum
as well as experiencing an effective potential.  To avoid this
momentum diffusion, we introduce a periodicity in the detuning rather
than in the intensity of the field.  We add a second laser beam in a
standing wave, which shifts one of the atomic levels by the
Autler-Townes effect \cite{CTbookTwo,origATref}.

Our atom is a three level atom with levels $|a\rangle$, $|b\rangle$ and
$|c\rangle$ as shown in Fig.~1.  There is a laser beam highly blue
detuned from the $|a\rangle~\leftrightarrow ~|b\rangle$ transition with
detuning $\Delta$ and frequency $\omega_{l}$ as shown, and a second
laser beam blue detuned from the $|b\rangle~\leftrightarrow ~|c\rangle$
transition with a frequency $\omega_{l2}$ and a much smaller detuning
$\delta$.  The second laser beam, which is closer to resonance, mixes
levels $|b\rangle$ and $|c\rangle$ into the dressed states $|1\rangle$ and
$|2\rangle$, as shown in Fig.~2.  When the Rabi frequency of the second
laser is much larger than its detuning, these states are approximately
given by \cite{Meystre}

\begin{mathletters}
	\begin{eqnarray}
		|1\rangle & = & \frac{1}{\sqrt{2}}\left(|b\rangle+|c\rangle\right)
		\label{dress1} \\
		|2\rangle & = & \frac{1}{\sqrt{2}}\left(|b\rangle-|c\rangle\right).
		\label{dress2}
	\end{eqnarray}
\end{mathletters}

The first laser is now detuned by an amount $\delta_j$ from the dressed
state $|j\rangle$, as shown in Fig.~2.  These detunings are given by

\begin{mathletters}
	\begin{eqnarray} \label{eq:detunings}
		\delta_1 = \Delta -\frac{\sqrt{\Omega({\bf r})^2+\delta^2}
				-\delta}{2}\\
		\delta_2 = \Delta +\frac{\sqrt{\Omega({\bf r})^2+\delta^2}
				+\delta}{2},
	\end{eqnarray}
where $\Omega({\bf r})$ is the resonant Rabi frequency of the second
laser interaction.
\end{mathletters}

The second laser is chosen so that it has a frequency which is so far
detuned from any transition from the ground state $|a\rangle$ that it
doesn't interact with it.  The frequency of the confining laser is
chosen so that it is too far off resonance to interact with the
$|c\rangle~\leftrightarrow ~|b\rangle$ transition.  The interaction
between the second laser and the atom is much stronger than that of
the first laser because it is stronger and much closer to resonance,
so we use the dressed state picture and treat the interaction of the
first laser semiclassically as a perturbative, electric dipole
interaction.  This means that the coherent motion of a single atom can
be found from the Hamiltonian:

	\begin{eqnarray} \label{eq:HamFirst}
	    \nonumber
	    \hat{H} = \frac{{\bf \hat{p}}^2}{2M}\;+\;\hbar(\omega_l-\delta_1)
		|1\rangle\langle 1|\; +\; \hbar(\omega_l-\delta_2)
		|2\rangle\langle 2|\\
		\rule{10mm}{0mm} +\; d E({\bf \hat{r}},t)
		\left(|b\rangle\langle a|+h.c.\right)
	\end{eqnarray}
where ${\bf \hat{p}}$ is the momentum operator, $M$ is the mass of the
atom, $\omega_l$ is the frequency of the first laser, $d$ is the
dipole moment of the $|a\rangle~\leftrightarrow ~|b\rangle$ transition,
and $E({\bf \hat{r}},t)$ is the electric field due to the first laser.

If $E({\bf \hat{r}},t)$ represents a beam propagating in the +z direction
then it will be of the form

	\begin{equation} \label{eq:Efield}
	       E({\bf \hat{r}},t) = \frac{C(\hat{x},\hat{y})}{2}
		\left(e^{-i(\omega_l t - k\hat{z})} + e^{i(\omega_l t
		- k\hat{z})}\right)
	\end{equation}
where $C(\hat{x},\hat{y})$ is the transverse profile of the electric
field.  We expand $|b\rangle$ and $|c\rangle$ into the dressed states
$|1\rangle$ and $|2\rangle$ and enter an interaction picture with
$\hat{H}_o=\hbar(\omega_l |1\rangle\langle 1| + (\omega_l -
\delta_2)|2\rangle\langle 2|)$.  We then ignore terms like
$|2\rangle \langle a|e^{-i\delta_{2}t}$ and
$|a\rangle\langle 1|e^{-i2\omega_{l}t}$ because $\delta_{2}$ and
$\omega_{l}$ are assumed to be large enough that these terms rotate
too fast to effect the long term dynamics of the atoms.  After this
rotating wave approximation, the interaction Hamiltonian becomes:

	\begin{equation} \label{eq:HamSecond}
		\hat{H}_I=\frac{{\bf \hat{p}}^2}{2M}\; -\; \hbar\delta_1
		|1\rangle\langle 1|\; +\; \frac{dC}{2\sqrt{2}}
		\left(|1\rangle\langle a|\langle a|e^{ik\hat{z}} + h.c.\right).
	\end{equation}

In momentum space this interaction couples the states $|a\rangle |q\rangle$
and $|1\rangle |q+k\rangle$, where $|q\rangle$ is the z-momentum eigenstate
with momentum $\hbar q$.  The Schr\"{o}dinger equation for the atom is
therefore reduced to two coupled partial differential equations.  We
then adiabatically eliminate the excited state\cite{Hope}.  This is valid
provided the detuning energy $\hbar \Delta$ is sufficiently large
compared to the transverse, recoil, and doppler kinetic energies;
provided we are only interested in time scales longer than the inverse
of these detunings, and provided the population in level $|1\rangle$
is sufficiently small.  The wavefunction of the $|1\rangle |q+k\rangle$
state can then be written in terms of the wavefunction of the
$|a\rangle |q\rangle$ state.  This gives the equation of motion for the
ground state wavefunction of the atom $\Psi_a(x,y)$:

	\begin{equation} \label{eq:eom}
		i\hbar \partial_t \Psi_{a} = -\frac{\hbar^2}{2M}\left(
		  \partial_x^2+\partial_y^2\right)\Psi_a \;+\;
		  \frac{d^2 C^2}{8\hbar\delta_1} \Psi_a.
	\end{equation}

The last term is an effective potential.  The adiabatic elimination
requires that the excited state population be very small which is
equivalent to ensuring that $(\hbar d^{2}/4) (C/\delta_1)^{2}\ll 1$.
Reducing the population of the excited state is also equivalent to
reducing the spontaneous emission rate of the atom.  The potential
barrier is proportional to $C^2/\delta_1$, and can therefore be made
large independently by making the laser stronger and detuning further.

Substituting Eq.(\ref{eq:detunings}) for $\delta_{1}$ into
Eq.(\ref{eq:eom}), we see that the effective potential $V({\bf r})$
experienced by the atom is given by

	\begin{equation} \label{eq:effPot}
		V({\bf \hat{r}})=\frac{d^2 C(\hat{x},\hat{y})^2}{8\hbar\left(
		\Delta-\frac{1}{2}
		  (\sqrt{\Omega({\bf \hat{r}})^2+\delta^2}-\delta)\right)}
	\end{equation}

The hollow optical fiber becomes an atomic waveguide when it is
guiding a travelling wave laser beam which is blue detuned from a
single transition in the atoms \cite{Marksteiner,Ito}.  In this case
both fields only depend on the longitudinal (z) and radial (r)
coordinates.

The energy spectrum of atoms in the fiber can be calculated
numerically when the potential can be separated:
$V(z,r,\phi)=V^z(z)V^r(r)$.  We will assume this to be true from this
point.  This is an approximation which becomes increasingly valid as
the hole in the fiber becomes smaller, as we will now show.  As
previously noted, the second laser beam must have a frequency which is
significantly different to the confining laser beam so that it does
not interact with the confining transition.  In particular, the second
laser can have a much longer wavelength, and it will therefore have a
different spatial mode in the fiber.  When the diameter of the hole in
the fiber becomes less than the wavelength of the second laser beam,
the beam will have a much lower radial dependence.  When the radial
dependence of the intensity of the second laser beam can be ignored
then the potential given in Eq.(\ref{eq:effPot}) is separable.  In
reality, the second beam will reduce in intensity towards the center
of the fiber.  This will cause the confining laser to have a larger
detuning from the atoms in the center, so the effect of including this
in the model would be to cause the atoms to be located in a deeper
potential well with a lower spontaneous emission rate.  These effects
should improve the practicality of an experiment, and are therefore
not important for a proof of principle calculation.

\section{Atomic energy spectrum in one dimension} \label{sec:1D}

If an atom is considered to be a plane wave travelling in a single
dimension, and there is a sinusoidal potential in that dimension, then
the solution to the Schr\"{o}dinger equation is well
known\cite{Mott,Kittel}.  Bloch's theorem states that for a
Hamiltonian of the form

\begin{equation}
     \hat{H} = -\frac{\hat{p}^{2}}{2M} + V(\hat{x}),
\end{equation}
the eigenstates $\Psi_k(x)$ of an atom in any potential with period
$\tau$ in the direction $x$ must be of the form:

	\begin{equation} \label{eq:BTheorem}
		\Psi_{k}(x) = e^{ikx}\sum_{n=-\infty}
				^{\infty}A_{n}e^{-i2\pi nx/ \tau}
	\end{equation}
where $A_n$ are coefficients.

This leads to the eigenvalue equation:

	\begin{eqnarray} \label{eq:1DEvalue}
		\sum_{n=-\infty}^{\infty}[-k_{n}^{2}+\frac{2M}{\hbar
			^{2}}(E_{k}-V)]A_{n}e^{-i2\pi nx/ \tau} = 0,\\
		\nonumber
			\mbox{where } k_{n}^{2}=(k-2\pi n/\tau)^2.
	\end{eqnarray}

When the variation in the potential is much less than the kinetic
energy of the particle, then the potential term in the above equation
can be treated perturbatively, and terms with the product of V and
$A_{n\ne 0}$ can be neglected.  For a sinusoidal potential, it has
been shown\cite{Mott} that the energy spectrum is approximately
$E_{k}=\hbar^2 k^2/2M$ except for the region $k\approx \pm\pi/\tau$,
where there is a gap equal to the height of the modulation of the
potential.  This band gap causes interesting conductive properties in
solids.

The real potential is not one dimensional.  In three dimensional
waveguides, variations of the potential in the transverse direction
can alter or remove the gap in the band structure.  The one
dimensional model ignores all transverse modes.  There are
non-degenerate transverse modes in the three dimensional model, so
their spacing must be large enough so that the gaps in the energy
spectrum are not covered.  From consideration of a particle in a box,
where the energy spacing decreases as the size of the box increases,
this will mean that the fiber will have some maximum inner radius.

\section{Atomic energy spectrum in three dimensions} \label{sec:Calc}

The energy spectrum for the three dimensional waveguide will now be
calculated.  The time independent Schr\"{o}dinger equation in
cylindrical co-ordinates ($z,r,\phi$) is:

	\begin{eqnarray} \label{eq:3DSE}
	   \left[\left(\frac{\partial ^{2}}{\partial r^{2}}
		+\frac{1}{r}\frac{\partial }{\partial r}+\frac{1}{r^{2}}\frac{\partial
		^{2}}{\partial \phi^{2}}+\frac{\partial ^{2}}{\partial z^{2}}\right)
		\right.&\\
	\nonumber
	   \left.+\frac{2M}{\hbar^{2}}\left(E-V(z,r)\right)
					\right]&\Phi(z,r,\phi)=0,
	\end{eqnarray}
where $\Phi$ is the atomic eigenstate with energy E in the potential
$V(z,r) = V^z(z)V^r(r)$.  The absence of a $\phi$ component of the
potential allows the eigenstate to be separated: $\Phi(z,r,\phi) =
\Psi(z,r)\Theta(\phi)$.  Eq.(\ref{eq:3DSE}) can then be
separated into functionally independent sides which then give a
simple analytic solution for $\Theta(\phi)$, and an eigenvalue
equation for $\Psi(z,r)$:

	\begin{equation} \label{eq:PhiSolution}
	   \Theta(\phi) = e^{im\phi}
	\end{equation}
where m is an integer, and

	\begin{equation} \label{eq:ReducedSE}
	   \left[-\frac{\partial ^{2}}{\partial r^{2}}-\frac{1}{r}\frac{\partial }
	   {\partial r}-\frac{\partial ^{2}}{\partial z^{2}}+\frac{2M(V-E)}
	   {\hbar^{2}}+\frac{m^{2}}{r^{2}}\right]\Psi=0.
	\end{equation}

This shows that the rotational modes induce an effective radial
potential.  This equation cannot be further separated, as the
longitudinal ``ripple'' in the potential couples the radial and
longitudinal motion. With a non-zero potential this equation must be
solved numerically.  The details of this calculation are given in
appendix \ref{app:Calc}.

The light induced potential on the two level atoms is proportional to
the intensity of the electric field.  For fibers with hole radii
larger than a few wavelengths, the evanescent field decays
exponentially from the walls.  For fibers with sub-micron hole radii,
the radial dependance of the electric field must be calculated
numerically for each fiber.  This was done using the techniques
described in references \cite{Marksteiner} and \cite{OWGbible}.

Gravity was neglected in this model, as the small hole size implies
that the effect will be a weak perturbation.  An estimate of the size
of this perturbation is $\Delta E = Mg \Delta h$ which for Helium
confined to a $0.5\mu$m region is about three orders of magnitude
smaller than the recoil energy.  The effect of the Casimir-Polder force
is approximated by reducing the potential at the walls of the fiber.
Reference \cite{Marksteiner} gives an estimate that this force halves
the height of the potential barrier seen by the atoms as they try to
reach the surface of the glass.  This change in the potential will
increase the required laser power, but this system already assumes
that the atoms are well confined, so the results of this model will
not depend on an accurate description of this force.

The parameters chosen for the calculation were based on a commonly
used transition in metastable Helium ($2^{3}S-2^{3}P$) with a
wavelength of $1083$nm, but the qualitative results will not depend
strongly on the chosen transition, and it is quite possible that
experiments would be based on a different transition or a different
atom.  The calculations were made using a fiber with a step profile
refractive index with core and cladding refractive indices of $1.5$
and $1.497$ respectively.  The radial width of the core was $3 \mu$m.

The Helium atoms in the fiber exhibit a band gap in their energy
spectrum for sufficiently small hole radii.  Fig.~3 shows the energy
spectrum of atoms in a typical fiber with increasing hole radius.  The
solid lines show the allowable energies, and the disallowed energies
are indicated by a shading on the energy axis.  Each line designates a
band, and the disallowed energies are the band gaps.  For small hole
radii, it is clear that there is a band gap between the lowest bands,
which becomes smaller as the hole becomes larger and the intensity of
the light in the fiber gets lower.  The spacing of the tranverse
modes is clearly smaller than the spacing of the rotational modes, as
the first excited rotational mode $m=1$ (dashed line) is higher than the
excited transverse modes.  Only the first few $m=0$ modes have been
shown, as there is only a significant band gap between the lower
bands.

As the size of the hole in the fiber approaches the de Broglie
wavelength ($\lambda_{dB}=h/p$) of the recoil cooled atoms, the
transverse energy spacing reduces such that the gap in the energy
spectrum is closed.  Small perturbative coupling between the
transverse modes would allow atoms to be excited from the lowest band
if such transitions were energetically allowed.  These couplings may
arise from interatomic collisions, gravitational, electric or magnetic
fields or by imperfections in the fiber or beam.  This means that the
atoms will only be confined to a single band if there is a band gap.
This will only occur when the hole is smaller than some maximum hole
radius which is about $1.5 \mu$m for this particular system.
Fig.~3(c) shows the energy spectrum for a radius which is $1.0 \mu$m,
and the higher rotational mode has nearly covered the band gap.
Fibers have already been manufactured with hole diameters as small as
$2.0 \mu$m.

While the atoms are in the fiber, the modulation depth can be varied
by altering the relative intensties of the two counterpropagating
laser beams which cause the level shift.  This controls the shape of
the energy bands and the size of the band gap.  Fig.~4 shows the
effect of decreasing the modulation depth.  As the modulation
decreases, the lowest band becomes lower and has more range while the
energy gap becomes smaller.

The large modulations cause spontaneous emission to be increased due
to the reduction in detuning of the first laser beam, and in fact
spontaneous emission is the key difficulty with this device.
Spontaneous emission losses depend sensitively on the energy of the
atoms, the size and refractive index profile of the fiber, the
available power in the confining laser and its detuning from the
dressed state transition.  The interesting atoms are those with an
energy that places them in the lowest band.  Classically these atoms
cannot enter the field further than the point where the potential
equals their total energy.  This means that an over-estimate of the
spontaneous emission rate can be found by considering the spontaneous
emission rate of an atom sitting at that point of maximum classical
potential with the minimum detuning of the confining laser.

For a hole diameter of $1.5 \mu$m, the spontaneous emission rate is of
the order of $1$ Hz, given $3$ W of guided laser power.  This is much
larger than that which might be achieved in free space, as there is
quite a large field at the center of the fiber which is causing the
atoms to become excited even when they are at the minimum of the
potential.  As may be the case for systems discussed in the next
section, if there was no electric field at the center, the spontaneous
emission rate would be as low as $0.01$ Hz.  The recoil cooled atoms
would take several seconds to pass through a few centimeters of fiber,
so attaining this limit for the spontaneous emission rate may be
enough to produce a practical device.

\section{Alternative laser schemes} \label{sec:AltSys}

The purpose of the confining potential is to confine the atoms to a
sufficently small area so that the splitting of the transverse modes
is larger than the size of the lower bands.  This allows a band gap to
form due to the periodic potential, and the gap is not filled by
higher transverse modes.  It should therefore be possible to
generalize this work to other designs for the physical layout of the
confining laser fields.

An alternate design to the hollow optical fiber experiment is a
variation on the traditional optical molasses apparatus.  A standing
wave in each of the two transverse directions produces a square sided
``tunnel'' in the laser field down which the atom can be guided.  A
longitudinal laser beam can then be used to produce the periodic
potential.  A significant advantage of this system is that there would
be zero electric field at the center of the atomic waveguide.  This
should allow lower spontaneous emission losses, as described in the
previous section.  The lack of rotational symmetry in this system
produces a calculational problem, however, as the potential cannot be
separated.  This means that an accurate calculation of the atomic
energy spectrum would be difficult to do numerically.

Other free space configurations may be used.  In particular, it might
be possible to produce a ``donut'' laser mode which has a node in the
center but retains the rotational symmetry which is useful in making
accurate calculations.

\section{Applications and Discussion} \label{sec:Appl}

The existence of a band gap in the energy spectrum for the atoms will
naturally suppress the excitation of atoms beyond the upper limit of
the band.  This means that atoms will tend to travel along the fiber
with less heating from incoherent sources.  Spontaneous emission
losses will dominate the loss of atoms from the system, and these
cannot be removed without using arbitrarily large laser
powers\cite{Hope}, so it is not feasible to construct arbitrarily long
waveguides.  The most important feature of the band structure in the
fiber is that it can be altered externally by changing the second
laser beam.  This means that interesting behaviour can be produced by
changing the band structure while atoms are in transit through the
waveguide.

The lowest band gap appears when the wavelength of the atoms is that
of the second laser beam.  This is illustrated in Fig.~4(c), where the
gap between the first two bands occurs at $k/k_{o}=1$ .  This means
that a beam of atoms which are cooled close to the recoil limit will
have a large population in the lowest energy band.  Since the atoms
are moving at speeds of the order of $3$ mm/s, an atomic beam would
take several seconds to pass through a centimeter of fiber.  This
means that the band structures can be changed so slowly that no
non-adiabatic heating effects need to be considered.

It can be seen from Fig.~4 that increasing modulation depths reduce
the width of the lowest energy band at the same time as increasing the
band gap.  If the modulation of the second laser was slowly increased
while there were atoms in that band, then the atoms would have a lower
energy spread.  This means that while atoms are guided through the
fiber, their energy spectrum can be altered in such a way that they
will be cooled.  This process is not expected to be necessarily
competitive with other cooling processes, but it is an example of
using the dynamic nature of the band structure in this fiber.  A band
structure with gaps in the energy spectrum which can be controlled
externally is a novel situation, and may lead to other interesting
effects.  Our cooling mechanism is similar to an effective three
dimensional cooling system which takes atoms in very deep potential
wells and adiabatically reduces the depth of the wells until they are
in free space \cite{Kastberg}.  This other adiabatic cooling system
may be an effective method of populating the lowest energy band.

The major limitation of this system is the spontaneous emission of the
atom from the excited state.  This can be made arbitrarily small by
increasing the detuning and the laser power, but if the laser is
detuned too much then it will interact with other levels in the atom.
For atoms cooled close to the recoil limit, the spontaneous emission
is limited by the maximum allowable detuning.

As spontaneous emission losses are likely to be quite high unless a
very low frequency transition can be used, there is some advantage to
using transitions based on metastable lower states.  Detectors of
these atoms rely on their large excitation energy.  If these atoms
spontaneously emit and do not return to the metastable state, then
they will not be detected, so the atoms that did not experience the
correct potential will have a reduced signal.  This alleviates the
problem of the high spontaneous emission rate experienced by the
atoms.

There are also possible methods for reducing the spontaneous emission
rather than simply eliminating the incoherent signal.  An example of
such a modification might be to use the potential produced by a Raman
transition instead of a single photon transition\cite{Hope}.  When
extremely high laser powers can be used to produce very low potential
barriers of the order of the recoil energy of the atoms, the
spontaneous emission rate is limited by the fact that the allowable
detuning has a maximum value beyond which the confining laser will
interact with different transitions.  This may be the situation for
the free space configuration described in the previous section.  In
this case the Raman scheme has been shown to allow reductions of the
spontaneous emission rate by removing this constraint.

\section{Conclusions} \label{sec:Concl}

This paper has shown that the energy spectrum for atoms guided through
a hollow optical fiber has at least two distinct bands separated by an
energy gap if the radius of the hole in the fiber is less than $1.5
\mu$m.  The size of this band gap and the energy range of the
lowest band may be controlled externally by altering the amount of
modulation of the second laser beam.

The ability to control the size of the band gap dynamically while the
atoms are in the system leads to many new possibilities.  This
situation does not exist in solid state physics where the band
structure is fixed.  For example, slowly increasing
the band gap while the atoms are being guided by the fiber will result
in a ``squeezing'' of the lowest energy band, so if the atoms were
originally cool enough to have a significant population in this band
then they will be further cooled.  The major limitation of this system
is that there would be a loss due to spontaneous emission.  Further work
may be useful in determining whether a similar system using lasers in
free space can produce similar or perhaps better results.

\acknowledgements

We would like to thank Dr M.Andrews for his useful discussions.

\appendix

\section{Details of Band Structure Calculation} \label{app:Calc}

In section \ref{sec:Calc} the Schr\"{o}dinger equation was shown to
reduce to the two dimensional equation:

	\begin{equation} \label{eq:SEagain}
	   \left[-\frac{\partial ^{2}}{\partial r^{2}}-\frac{1}{r}
	   \frac{\partial}{\partial r}+\frac{m^{2}}{r^{2}}-\frac{\partial ^{2}}
	   		{\partial z^{2}}+\frac{2M(V-E)}{\hbar^{2}}\right]\Psi=0.
	   \end{equation}

In the full three dimensional case, shown in Eq.(\ref{eq:3DSE}), the
longitudinal co-ordinate satisfied the conditions for Bloch's theorem.
The shift to cylindrical coordinates and solution of one of the
degrees of freedom has not changed that fact, so we make the
substitution:

	\begin{equation} \label{eq:Substn}
		\Psi_{k}(r,z) = e^{ikz}\sum_{l=-\infty}
				^{\infty}A_{l}(r)e^{-i4\pi lz/ \lambda}
		\end{equation}

This leads to the eigenvalue equation:

	\begin{eqnarray} \label{eq:ContRSE}
		\sum_{l}\left[-\frac{\partial ^{2}}{\partial r^{2}}-
			\frac{1}{r}\frac{\partial}{\partial r}+\frac{m^{2}}
	   		{r^{2}}+k_{l}^{2}+\frac{2MV}{\hbar
			^{2}}\right]A_{l}e^{ik_{l}z}\\
		\nonumber
			= \frac{2ME}{\hbar^{2}}\sum_{l}A_{l}e^{ik_{l}z},
	\end{eqnarray}
where $k_{l}^{2}=(k-4\pi l/\lambda)^2$.  The potential is
multiplicatively separable into longitudinal and radial components,
$V(z,r)=V^z(z)V^r(r)$.

To transform radial wavefunctions into Fourier space, we consider
Eq(\ref{eq:SEagain}) and replace $\Psi(z,r)$ with $\Phi(z,r)=r^{1/2}
\Psi(z,r)$.  This leads to the Schr\"{o}dinger equation:

	\begin{equation} \label{eq:SEinPhi}
	   -\frac{\partial ^{2}\Phi}{\partial r^{2}}-\frac{\partial
	   ^{2}\Phi}{\partial z^{2}}+\left(\frac{m^{2}-
	   \frac{1}{4}}{r^{2}}+\frac{2M(V-E)}{\hbar^{2}}\right)\Phi=0.
	   \end{equation}

We then produce the final version of the Schr\"{o}dinger equation:

	\begin{eqnarray}
		\sum_{l'}\sum_{n'}{\Huge [}\left(\textstyle k_{l'}^{2}+ \left(
		\frac{2 \pi n'}{R}\right)^2\right)\delta_{l,l'}\delta_{j,j'}&
		\\
		\nonumber
		+\frac{\delta^{l,l'}}{R}\int_{0}^{R}\frac{m^{2}-\frac{1}
		{4}}{r^{2}}&\!\! e^{i2\pi (n-n')r/R}dr \\
		\nonumber
		\rule{20mm}{0mm} +V_{l-l'}^z V^r_{n-n'}{\Huge ]}{\cal A}_{l',n'}& =
		\displaystyle\frac{2ME}{\hbar^2} {\cal A}_{l,n}
	\end{eqnarray}
where
	\begin{equation}
		V^z_{l}=\frac{2}{\lambda}\int_{\frac{\lambda}{4}}^{\frac{\lambda}{4}}
	\!\!V^z(z) \frac{2M}{\hbar^{2}} e^{4 \pi i l z/\lambda} dz
	\end{equation}
and
	\begin{equation}
		V^r_{n}=\frac{1}{R}\int_{0}^{R}\!\!V^r(r)e^{2 \pi i n r/R}dr
	\end{equation}
and
	\begin{equation}
	    \Phi_{k}(r,z) = \sum_{l=-\infty}^{\infty}\sum_{n=-\infty}^{\infty}
	    A_{l,n}e^{ik_{l} z}e^{-i2\pi nr/ R}.
	\end{equation}

This is now in a suitable form to perform a numerical calculation of
the atomic energy spectrum.

This seems an unusual numerical method for solving the Schrodinger
equation in cylindrical coordinates, as the most obvious
transformation to make on Eq.(\ref{eq:ContRSE}) is the Hankel transform,
which expands the radial wavefunctions in terms of Bessel functions
and removes the singularity.  If we do this using the notation ${\cal
A}_{l}^{m}(\kappa) = \int_{0}^{\infty}dr \;r A_{l}(r) J_{m}(\kappa r)$
we can show that:

	\begin{eqnarray}
		\int_{0}^{\infty}\!\!\!\! d\kappa'\sum_{l'}\left\{
		  (\kappa'^{2}+k_{l'}^2)\delta_{l,l'}\delta(\kappa
			-\kappa')\right.&\\
		\nonumber
		\left.+V_{l-l'}^z\kappa' V^r(\kappa,\kappa')
			\right\}{\cal A}_{l'}^{m}(\kappa')
		=&\displaystyle\frac{2ME}{\hbar^2}{\cal A}_{l}^{m}(\kappa)
	\end{eqnarray}
where
	\begin{equation}
		V^r(\kappa,\kappa')=\int V^r(r)rJ_{m}(\kappa r)J_{m}
						(\kappa' r)dr
	\end{equation}
and
	\begin{equation}
		V^z_l=\frac{2M}{\hbar^2}\frac{2}{\lambda}\int_
		    {-\frac{\lambda}{4}}^{\frac{\lambda}{4}}V^z(z)
		    e^{i4\pi lz/\lambda}dz
	\end{equation}

For the purpose of producing a numerical calculation, the indices on
the wavefunction components must be reduced to a finite set.  Placing
the Hankel coordinate $\kappa$ on a grid is unsatisfactory, as there
is very slow convergence for the integral defining
$V^{r}(\kappa,\kappa')$, and it converges very slowly for large
$\kappa$.  This is why a better numerical performance can be achieved
by transforming Eq.(\ref{eq:ContRSE}) into Fourier space.

\begin{figure}
 \label{fig:lasers}
 \caption{Atomic states and laser configurations}
\end{figure}

\begin{figure}
 \label{fig:dressed}
 \caption{Dressed states}
\end{figure}

\begin{figure}
\label{fig:resultsHoleRadius}
\caption{The energy spectrum of atoms in a hollow fiber
with increasing hole radius.  The horizontal axis is quasimomentum
$\hbar k$ in units of the momentum of a single photon $\hbar k_{o}$
from the confining laser.  Energy is measured in units of the recoil
energy E$_{R}=(\hbar k)^{2}/(2M)$of a single atom.  Note that the
energy scale is different in each figure, as the atoms are immersed in
a more intense light field when the fiber has a smaller hole.  The
dashed lines show the higher rotational mode, m=1, and the solid lines
indicate the lowest rotational mode, m=0.  Higher transverse modes in
the m=0 spectrum are not shown in (b) and (c), for clarity.
(a),~(b)~and~(c) have hole radii of $0.25\mu$m, $0.5 \mu$m, and $1.0
\mu$m respectively.  These figures have a modulation of
$M=1-(\mbox{min}(V^{z})/\mbox{max}(V^{z}))=0.02$, as it was in
Fig.~4(b).}
\end{figure}

\begin{figure}
\label{fig:resultsMod}
\caption{The energy spectrum for atoms in a hollow fiber
with increasing modulation of the potential.  The horizontal axis is
quasimomentum $\hbar k$ in units of the momentum of a single photon
$\hbar k_{o}$ from the confining laser.  Energy is measured in units
of the recoil energy E$_{R}=(\hbar k)^{2}/(2M)$of a single atom.  The
modulation is measured by the parameter
$M=1-(\mbox{min}(V^{z})/\mbox{max}(V^{z}))$ where $V^{z}$ is the
z-dependence of the confining potential.  (a),~(b)~and~(c) have values
for M of 0.04, 0.02 and 0.005 respectively.  The hole radius for all
three figures was $0.5 \mu$m, as it was in Fig.~3(b).}
\end{figure}

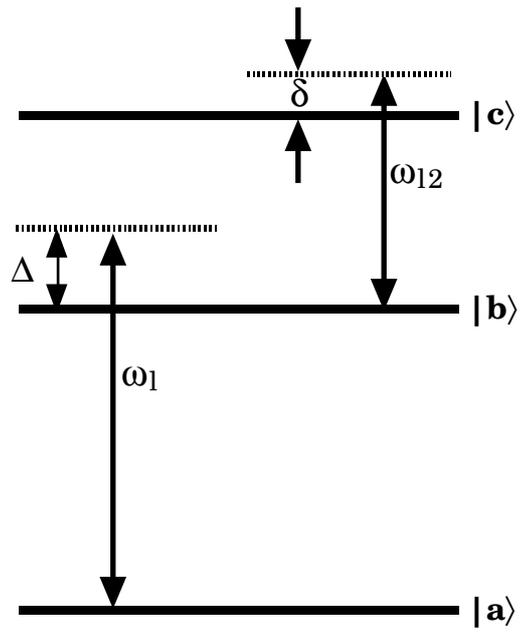
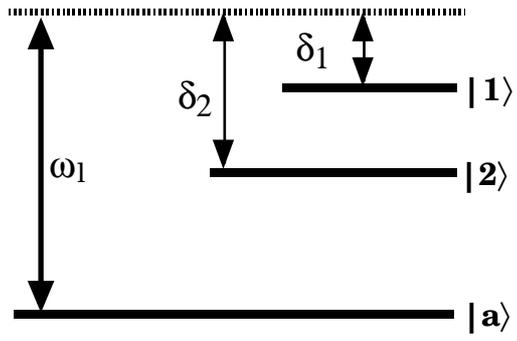
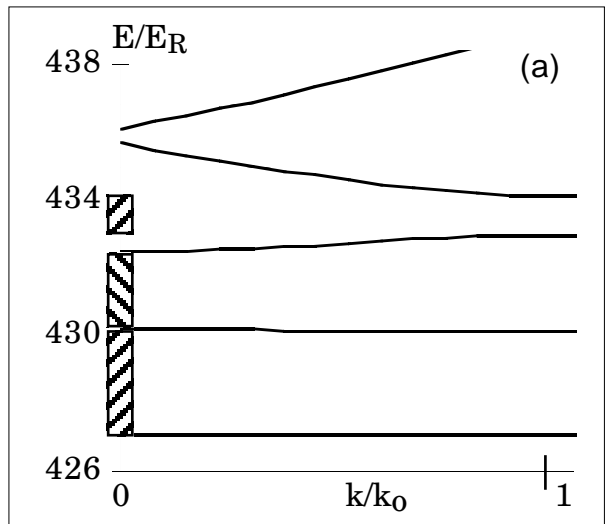
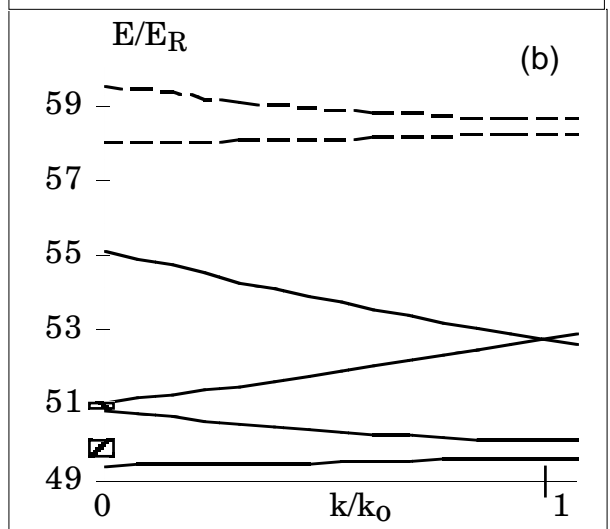
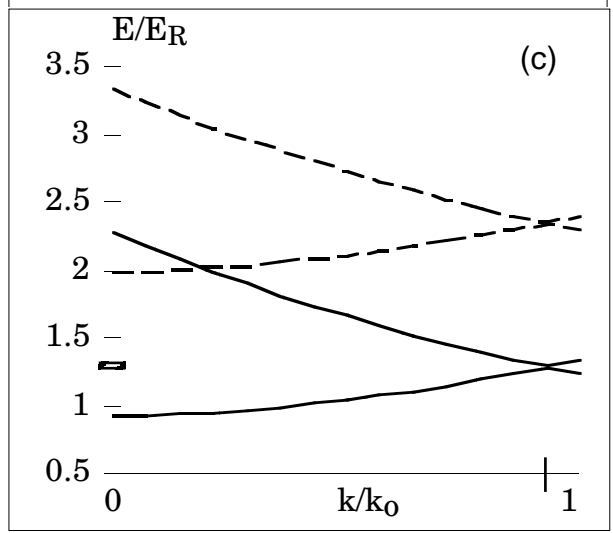
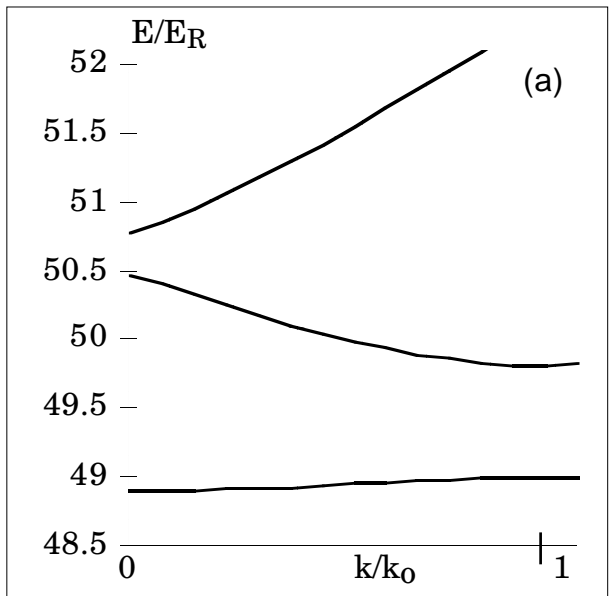
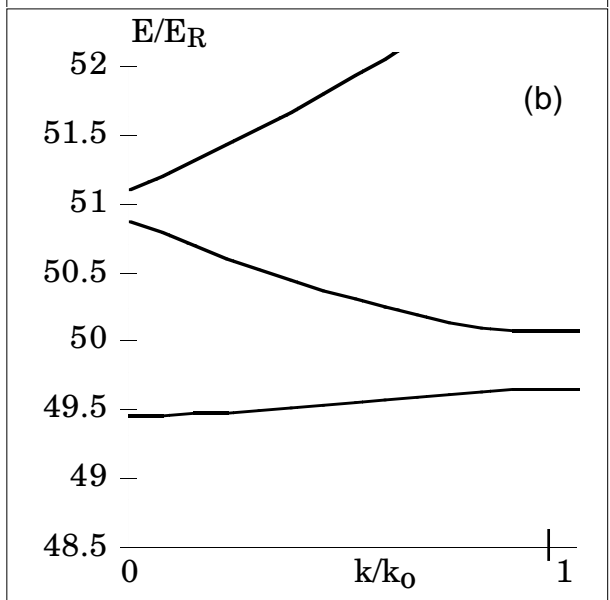
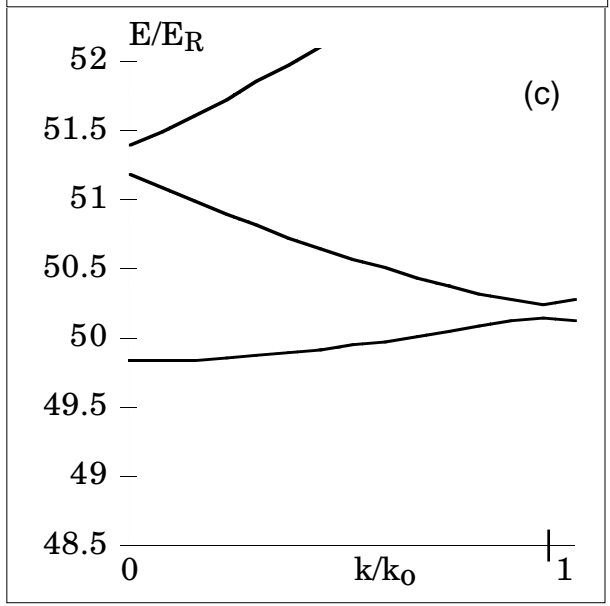
\end{document}